# Reversed Photoeffect in Transparent Graphene Nanocapacitors

A. Belkin, E. Ilin, I. Burkova, and A. Bezryadin
Department of Physics, University of Illinois at Urbana-Champaign, Urbana, IL 61801

ABSTRACT: *Electronic properties of ultrathin dielectric films consistently attract much attention since they play important roles in various electronic devices, such as field effect transistors and memory elements. Insulating properties of the gate oxide in transistors represent the key factor limiting Moore's law. The dielectric strength of the insulating film limits how much energy can be stored in nanocapacitors. The origin of the electric current in the nanometer-scale insulating barrier remains unexplained. Here we present an optically transparent Al-$Al_2O_3$-graphene nanocapacitor suitable for studying electronic transport in calibrated nanoscale dielectric films under high electric fields and with light exposure. The controllable flow of photons provides an additional powerful probe helping to resolve the puzzle of the electric conductivity in these high-quality insulating films. The dielectric alumina, $Al_2O_3$, is deposited by atomic layer deposition technology. With this device we observe a photon-assisted field emission effect, in which the effective barrier height is reduced by a quantity equal to the photon energy. Our main finding is a reversed photoeffect. Namely, at sufficiently high bias voltages the current through the dielectric film decreases as the light intensity increases. Moreover, higher photon energies correlate with stronger decreases of the current. To explain this reversed photoeffect, we present a qualitative model based on a conjecture that electrons leak into the dielectric and form charged sandpile-like branching patterns, which facilitate transport, and which can be dispersed by light.*

INTRODUCTION

Ideally, dielectrics are supposed to carry no current if the voltage is less than the breakdown voltage. Yet, leakage currents occur at voltages much lower than the critical level at which the dielectric breakdown occurs. Leakage currents in gate oxides of common electronic circuits represent the key limiting factor slowing down the Moore's law progression to ever smaller and denser electronic circuits. Regardless of its importance in a great variety of nanoscale electronic devices, the physical origin of the leakage current in common dielectric is not well understood, especially at the nanoscale[1,2,3].

      Graphene nanocapacitors[4], while exhibiting interesting quantum effects[5], provide a testing ground for studying the dielectric conductivity and leakage currents. Such capacitors may have a practical importance also, since they have been considered as possible candidates for energy storage characterized by durability, high power density and efficient operation at various temperatures starting from cryogenic temperatures and up to the melting point of the composing materials[6,7,8].

An interesting recent development in the field of graphene capacitors is the observation[5] of anomalous size-dependent increase of capacitance in capacitors with boron nitride as the dielectric layer. The original explanation of the phenomenon was given in terms of a negative quantum capacitance. However, more recent theory argues that the anomalous capacitance of such Van der Waals device can have a purely electrostatic origin. The explanation is developed based on the dependence of the relative permittivity on the film thickness[9].

Apart from nanocapacitors, such oxides are used in embedded random access memory[10], complementary metal-oxide-semiconductor transistors[10], and radio-frequency capacitors[11]. A simplistic view on the dielectric layer as an energy barrier for the electrons suggests that the leakage in capacitors is solely due to the direct tunneling current. Yet, in fact, there is a realm of other mechanisms for the dielectric conductivity, which have been suggested but not well understood so far[12].

Here we report experiments on transparent nanocapacitors. They allow a direct light exposure of all three layers of the nanocapacitor. Such option allows a detailed study of the photon influence on the current in the dielectric under strong voltage bias. Our key result is the observation of a reversed photoeffect, in which light exposure leads to a reduction of the current through the insulator. To explain this, we employ results of previous work[13] demonstrating that electrons can form a branching pattern as they penetrate to a low-conductance materials. Within one branch electronic motion is quasi-one-dimensional. One dimensional penetration of charges into a disordered medium is governed by the theory sandpile-style theory of the penetrating charge[14]. Generally speaking, such charge penetration sets the insulating barrier performance limit in the nanocapacitors.

RESULTS

Our devices are metal-insulator-graphene capacitors fabricated on glass substrates. The bottom layer is 10 nm aluminum film deposited by mean of electron-beam evaporation in a chamber with the base pressure ~$10^{-9}$ Torr (AJA). After the aluminum deposition the samples were quickly transferred (in air) into an atomic layer deposition (ALD) system where the whole surface was homogeneously covered by alumina using trimethylaluminum (TMA) and water vapor precursors at 80°C. The time of exposure of Al film to the atmospheric oxygen during sample transfer from Al to $Al_2O_3$ deposition chambers did not exceed 10 minutes, thus the thickness of naturally grown oxide is approximately 2 nm[15,16]. The thicknesses of the tested ALD-grown alumina are 10, 13 and 15 nm. The last step of the sample fabrication process was the deposition of a CVD (chemical vapor deposition) graphene (Trivial Transfer Graphene from ACS Material). It was transferred on the surface using the standard procedure provided by the manufacturer. The typical capacitor area is $A=1$ mm$^2$ = $10^{-6}$ m$^2$.

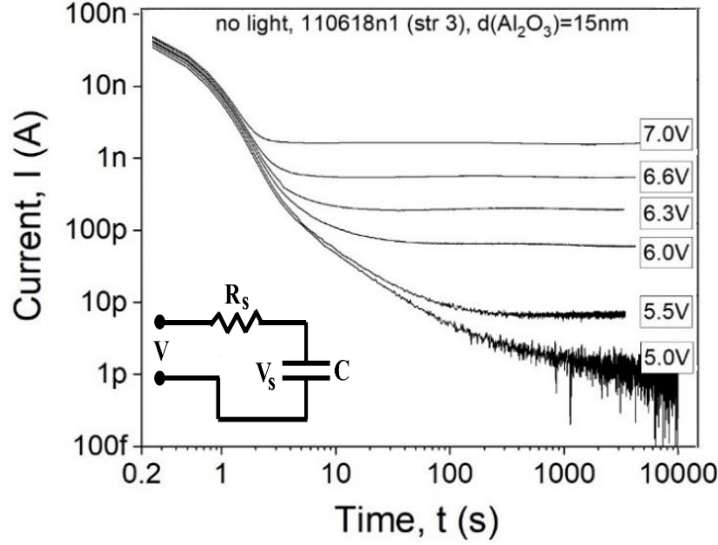

*Figure 1. Current through a representative nanocapacitor as a function of time at different applied voltages (V). The thickness of the alumina is 15 nm. The parameter is the applied voltage, which is applied to the capacitor connecter in series with the resistor $R_s$ =100 MΩ. These measurements have been performed in an optically non-transparent, "dark", Faraday cage. The insert shows how the voltage is applied through a series resistor.*

    The experimental circuit, similar to Ref.4, used in all reported experiments contains three main elements connected in series, namely a nanocapacitor, a series resistors $R_s$, a voltage source (Fig.1, insert). The current in the circuit is measured using an ammeter Keithley 6517B, which provides a sub-pA current resolution. This device is also equipped with a voltage source, $V$. The standard resistor, $R_s$ =100 MΩ, serves to limit the current in case of a breakdown. The sample was placed into a metal enclosure (Faraday cage), which helped to shield the nanocapacitor from external electromagnetic noise. Shielding is essential if the measured currents are expected to be low, such as in the pA and sub-pA ranges for example. Our Faraday cage, which screens the sample from light as well as any other external electromagnetic noise, was equipped with a photodiode providing an influx of monochromatic photons at the device surface. The measurements were controlled using LabView software.

    The electrical current in the circuit described above consists of three main components: (1) the capacitor plates charging current (CPCC), (2) the dielectric charging current, and (3) the leakage current[4]. The first two mechanisms are transient while the last one continues indefinitely long. The CPCC characteristic time scale, $\tau$, is determined by the capacitance, $C$, and the resistor inserted in the charging circuit, $R_s$. The capacitance of our samples was ~4 nF and the series resistor was $R_s$ = 100 MΩ, thus $\tau = R_s C \approx 0.4$ s. Since the charging current drops exponentially, its contribution to the total current becomes negligible after a few seconds (Fig.1). On the contrary, the dielectric-charging current continues for hundreds of seconds[4] (Fig.1). Thus, to detect the true leakage current we need to wait, sometimes for a long time, until the current reading saturates. The waiting time, needed to measure the true leakage current depends on the magnitude of the leakage current, which is defined by the applied voltage. For example, at $V = 5.5$ V, the dielectric charging current is dominant at 4 s $< t <$ 200 s. The leakage current is of the order of 10 pA, so the waiting

time until the saturation begins is of the order of ~300 s. On the other hand, at a higher voltage, e.g., $V = 7$ V, the leakage current is larger than 1 nA. Since the dielectric charging current is typically much weaker than this value, the leakage current can be measured just after a few seconds after the voltage is applied or changed.

A general conclusion from the discussion above is that to measure the leakage current correctly it is usually necessary to measure the time dependence of the current in the circuit, after a fixed voltage is applied. There are in fact two different transient processes which contribute to the total current. They can be mistakenly taken as the leakage. If the leakage is small, the waiting time, needed to reach the saturation current, can be very long, possibly hundreds or even thousands of seconds (as in the case $V = 5$ V). At $V < 5$ V the leakage current is undetectable with this particular thickness of the dielectric layer (15 nm), corresponding to 0.33 GV/m (at $V = 5$ V).

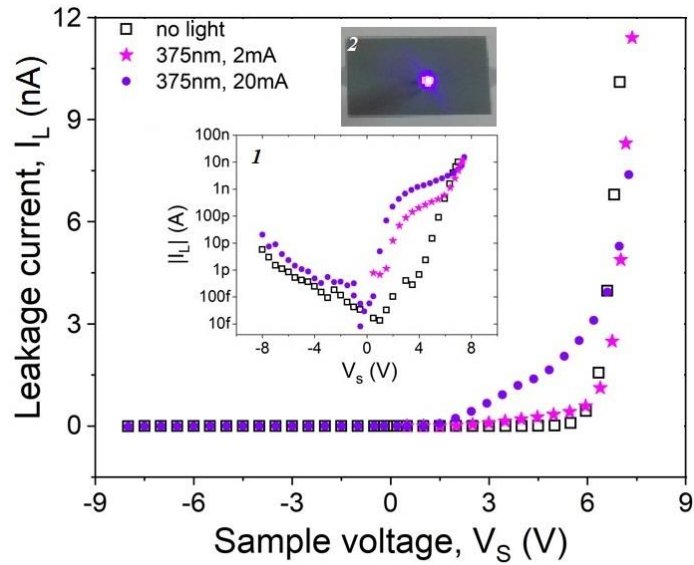

*Figure 2. True leakage current versus the voltage on the sample 110618n1 (the nanocapacitor). Dark state measurements are shown by open black squares. Positive $V_S$ corresponds to the positive voltage on the graphene plate. The thickness of the alumina is 15 nm. The Inset 1 presents the amplitude of the leakage current on the log scale. Violet circles: the steady state leakage current under a light exposure. Magenta stars: same as violet circles, but the light intensity is about 10 times lower. To be more precise, the current in the photodiode was reduced by a factor of ten in the experiments shown by magenta stars as compared to the measurements presented by violet circles. The Inset 2 shows a demo photo of the transparent $Al/Al_2O_3/Al$ nanocapacitor: The thicknesses of each metal plate is 5 nm and the thickness of the alumina is 10 nm. A blue photodiode was placed behind the capacitor and it is well visible through the capacitor, illustrating the fact that the nanocapacitor is optically transparent.*

From a series of $I(t)$ measurements similar to those shown in Fig. 1 we extract values of leakage currents, $I_L$, at different sample voltages, $V_S$. The results are given in Fig.2. Each data point here was obtained from a horizontal linear fit of the tail of the $I(t)$ curve, where all time-dependent transient processes decayed (Fig. 1). The sample voltage is calculated by the formula $V_S = V - IR_S$, where $R_S$ is the series resistor, $I$ is the measured current in the circuit and $V$ is the applied voltage.

This formula takes into account the fact that the voltage *V* is applied not directly to the capacitor, but to the capacitor and the resistor $R_S$ connected in series. As it follows from Fig. 2, the dependence $I_L(V_S)$ is asymmetric with respect to $V_S=0$ axis. The "high" voltage terminal is always connected to the graphene plate of the capacitor, which means that positive *Vs* is equivalent to the positive voltage on the graphene plate of the capacitor, and $Vs < 0$ indicates that the graphene plate is at a negative potential. The dependence of the leakage current on the voltage polarity can be attributed to different values of the work functions of aluminum and graphene. The energy required to remove an electron from the Al cathode ranges from ~3 eV to ~4 eV. For the graphene cathode this energy is significantly higher, reaching 4.9-5.2 eV [17]. Also, the asymmetry might be due to the fact that the surface of graphene is atomically flat as compared to non-epitaxially grown Al films, which have a roughness typically on a scale of at least a few nanometers. The flatter graphene surface impedes electric field amplification processes, typically occurring due to sharp features present on the surface of a metal electrodes. This circumstance makes graphene superior to Al in capacitors where a low leakage is required. We assume that the current in the dielectric is carried by the electrons, dispatching from the negatively charged plate of the capacitor.

Below we discuss the impact of light exposure on the leakage current. We will argue that the mechanism for the leakage current is a field emission process based on the phenomenon of quantum tunneling through a potential barrier, defined by the work function of the material of the capacitor plates. The photons provide additional energy to the electrons. Thus, the effective barrier is reduced by the energy equal to that of the single photon. Such effect can be classified as photon-assisted field emission.

Also, since quantum tunneling depends exponentially on the electric field, the electronic branching "sandpile" structure (EBSS), formed in the dielectric due to the bulk charge penetration, leads to an increased field current, i.e., the leakage current. Since the leakage current slowly increases in time, it will be argued that the EBSS growth in time. We will present arguments that light photons can destroy EBSS and therefore can reduce the leakage, producing the reversed photoeffect.

In order to study the influence of light on the leakage current we used light emitting diodes (LED) mounted above the surface of the nanocapacitors. The maximum power of the diodes was 38 mW/sr. The distance from the photodiode to the nanocapacitor was ~1 cm. We note that this power corresponds to an electric field intensity which is by a factor of ~$10^6$ lower than the electric field at the breakdown. Thus, this weak light source cannot lead to any structural damage within the dielectric.

Figure 2 demonstrates how a flux of photons impacts the leakage current in the nanocapacitors under investigation. In this case, the diode emits ultraviolet photons with wavelength 375 nm. When graphene is negatively biased the light causes only a small change in the current, which can be noticed if logarithmic scale is used (see Fig. 2, insert). Thus, in what follows we will focus on the experiments where a negative potential is applied to the Al plate, meaning $V_S>0$. Figure 2 shows that, for example for $V_S = 4$ V, the applied ultraviolet light can increase the leakage current by a factor of ~1000. If the current in the light-emitting photodiode is reduced by a factor 10, from 20 mA to 2 mA, (see the stars in Fig.2) then the leakage current exhibits a reduction by a factor ~10 also. This observation suggests that the light-induced leakage current is proportional to the number of photons hitting the capacitor per each second.

The transparency of the nanocapacitors is directly illustrated in Fig. 2 (insert 2). The observed photoeffect and the asymmetry are linked to the transparency of the graphene capacitor plate as well as individual transparencies of all the other layers. The majority of photons (~98%) travel through the graphene plate without being absorbed. Those 2% which get absorbed produce a much smaller number of high energy electrons that give relatively small contribution to the leakage current. The situation is very different when photons hit the Al electrode. Since this electrode is considerably thicker than the graphene layer, it absorbs more photons providing a larger number of excited electrons near the Al surface. These high-energy electrons have a higher probability to overcome the potential barrier associated with the insulator and reach the opposite electrode (graphene). Physically, the effect of photons is that they excite the electrons to higher energy levels and such exited electrons can tunnel through the surface energy barrier with an exponentially higher probability, thus making the leakage current at least three orders of magnitude larger (Fig.2).

Next, we investigate the photoeffect at different values of the wavelength of the photons by using photo-emitting diodes of different colors. The results are shown in Fig.3 in a log-linear format. At low bias voltages ($V_S < 5.5$ V), the current through the capacitor increases with the photon energy (the inverse wavelength). The strongest photo-current is observed in the case of ultraviolet light with a wavelength of $\lambda = 375$ nm and the corresponding photon energy $E_{ph} = h(c/\lambda) = 3.3$ eV. The leakage current becomes detectable at the $V_S \sim 0.5$ V. Here $h$ is the Planck's constant and $c$ is the speed of light. Since the current appears almost at zero bias when such photons present, we expect that the barrier for the electron tunneling through the capacitor is just slightly higher than the photon energy, but certainly not lower than 3.3 eV. This energy defines a lower bound and a rough estimate of the effective energy barrier for the electron tunneling, as will be discussed below.

The most important finding that we present here is the reversed photoeffect (Fig.3a), observed at higher voltages ($V_S > 5.5$ V). Indeed, at $V_S > 5.5$ V the current is the largest in the case of dark measurements (black squares) and the lowest in the case of ultraviolet irradiation (crosses). We suggest that at high bias voltages the electrons penetrate into the dielectric (alumina) to a certain distance, which is suggested by the theory of Ref. [14]. If the dielectric medium is disordered, the theory predicts that the electronic density should drops linearly with the increasing distance to the Al-Al$_2$O$_3$ interphase. In other words, injected electrons in the alumina form an electronic branching "sandpile" structure, EBSS, the edge of which marks the line up to which the electrons were able to creep at a given bias voltage. Thus, the effective thickness of the dielectric is reduced by the amount equal to the width of the electronic branching "sandpile" forming in the insulator of the capacitor. We conjecture that the reversed photoeffect occurs due to a partial or complete destruction of the EBSS structure by the flux of incoming photons. We note that EBSS structure might have a complex shape whereby electrons penetrate into the insulated in certain areas, possibly in a form of sharp tips. The electric field may be enhanced at such tips and the direct photoeffect may be amplified at those tips. According to this view, the destruction of the EBSS can produce a significant reduction of the photocurrent, as is indeed observed.

Experimentally, the reversed photoeffect is especially pronounced in the linear format plots of Fig.3b. For example, at $V$=6.5 V the current through the capacitor irradiated by ultraviolet photons (magenta crosses) is ~2.5 nA, while the current of the same capacitor measured in the dark is ~27 nA, which is about a ten-fold increase.

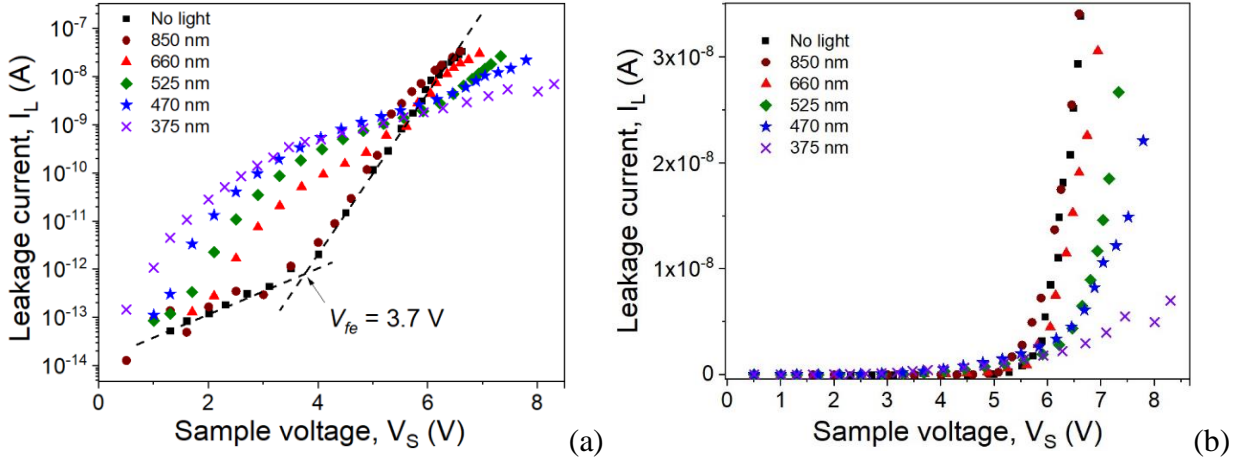

*Figure 3. (a) Voltage dependence of the leakage current in the presence of light exposure of the sample. The alumina thickness is 10 nm. Black squares correspond to the "no light" condition, circles – infrared light (850 nm), triangles – red light (660 nm), diamonds – green light (525 nm), stars – blue light (470 nm), crosses – ultraviolet light (375 nm). (b) Same data is in (a), but in the linear axis format. The reversed photoeffect is especially clear with this axis choice.*

Different mechanisms can be responsible for the current through the dielectric layer of the capacitor. At low voltages it is either electron diffusion or hopping. As the applied voltage increases and becomes larger than the energy barrier ($V_S > \phi_b$), electrons begin to tunnel from a capacitor plate into the conduction band of the dielectric. This field-emission process causes a huge increase of the leakage current. It is marked by a noticeable change of the slope of the log($I$) versus $V$ plot (Fig.3a). The critical voltage at which the field emission leakage begins is marked $V_{fe}$. We find that $V_{fe} = 3.7$ V. This voltage represents the effective barrier height defining the field emission process, $\phi_b \sim 3.7$ eV, which is also in agreement with the above estimate of a lower bound for the barrier, namely the 3.3 eV level.

The expression for the field-emission current density is[18]: $J = \alpha E^2/\phi_b \exp(-\beta \phi_b^{3/2}/E)$ (in units A/m$^2$), where $\alpha = e^3(m_0/m^*)/(8\pi h) \approx 1.54 \cdot 10^{-6}(m_0/m^*)$ (in units A eV/V$^2$) and $\beta = 8\pi(2m_0)^{1/2}/(3eh) \approx 6.83 \cdot 10^9 (m_0/m^*)^{-1/2}$ (in units V/m/eV$^{3/2}$), $\phi_b$ (in units eV) is metal-insulator barrier height, $m_0$ is the free electron mass, $m^*$ is effective electron mass in the insulator.

For the convenience of the analysis we will use voltage $V$, which is related to the electric field $E$ as $V=Ed$. Taking the typical dielectric thickness $d=10$ nm, the equation for the field-assisted tunneling become: $J = 6.7 \cdot 10^{10} V^2/\phi_b \cdot \exp(-33\phi_b^{3/2}/V)$ or

$$I/V_S^2 = 1.54 \cdot 10^{-6} \frac{m_0}{m^*} \frac{1}{Ad^2 \phi_b} \exp\left(-6.83 \cdot 10^9 \frac{(m^*)^{1/2} d}{m_0^{1/2}} \frac{\phi_b^{3/2}}{V_S}\right) \; A/m^2.$$

The total current is related to the current density as $I=AJ$, where the capacitor area is $A=1$ mm$^2 = 10^{-6}$ m$^2$.

According to the equations above, it is useful to plot the logarithm of the current divided by the voltage squared versus the inverse voltage ($1/V_S$). Such plots are expected to be linear. The results are shown in Fig.4, where the linear dependence is observed. The dashed lines show the best linear fits. Consider for example the measurement done in the dark (Fig. 4a). The expected linear dependence in this case is seen at voltages larger than ~3.7 V, i.e. at inverse voltages lower than 0.27 (in SI units).

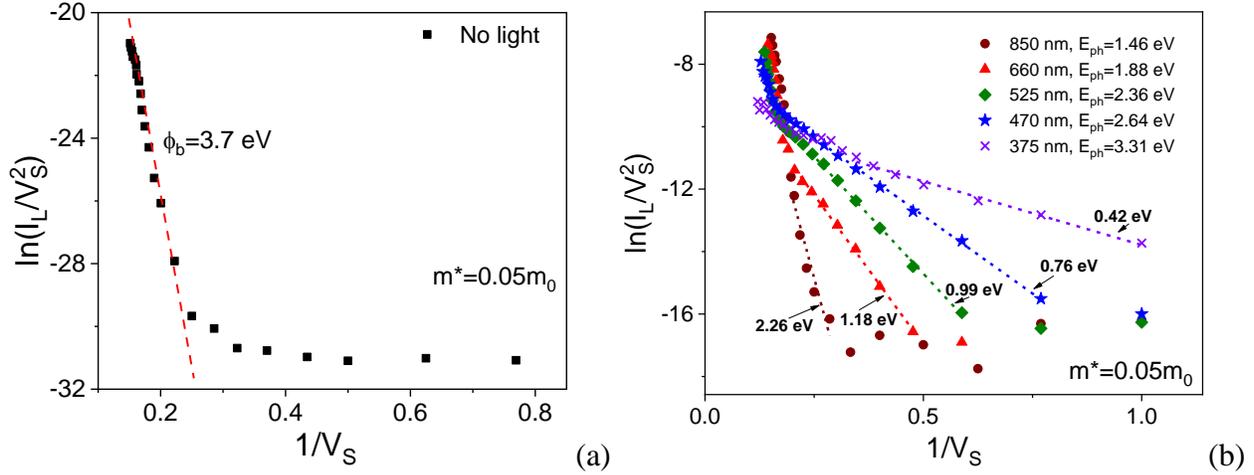

*Figure 4. Fowler-Nordheim (FL) plots of the data shown in Fig. 3: (a) The "no light" condition. (b) Measurements with various light colors. The circles – infrared light (850 nm), the triangles – red light (660 nm), the diamonds – green light (525 nm), the stars – blue light (470 nm), the crosses – ultraviolet light (375 nm). Slopes of the best fits (dashed lines) provide information about the effective tunnel barrier height for each photon energy.*

The slope of the linear fit can be used to find the effective energy barrier $\phi_b$. For example, the best fit in Fig.4a provides $\phi_b$=3.7 eV (for the "no light" case), using $m^*$=0.05$m_0$, reported previously[19]. The effective energy barriers for the curves measured under light exposure are indicated near each curve in Fig.4b. The effective energy barrier is reduced with the photons energy according to the Planck's formula $E_{ph}=hf$, connecting the frequency and the energy. Thus $\phi_b(f)=\phi_b(0)-E_{ph}=(3.57 \text{ eV})-E_{ph}$ (see Fig.5), since the photon energy is added to the energy of the electrons absorbing the photon. Thus, the effective tunnel barrier is reduced by the energy equal to a single photon energy. The value $\phi_b(0)$=3.57 eV is obtained from the fit of Fig.5 taking into account all the measurements with different photon colors (energies). It is close to the value estimated from the "no light" measurement of Fig.4a.

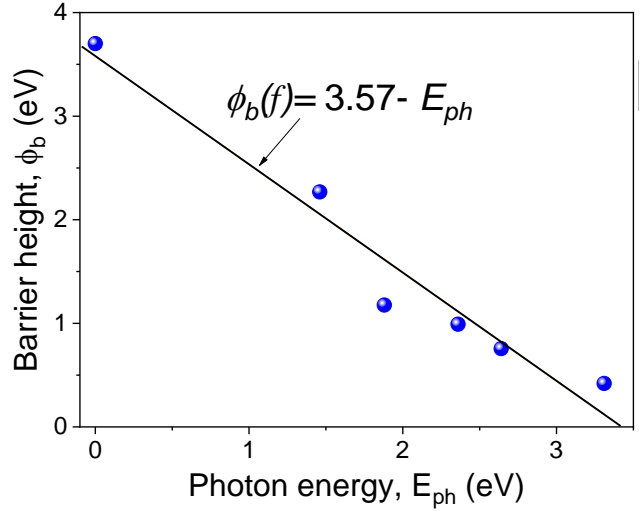

*Figure 5. The effective energy barrier as a function of the photon energy. The formula for the fit is $\phi_b(f)=3.57-E_{ph}$, where $E_{ph}$ is the photon energy.*

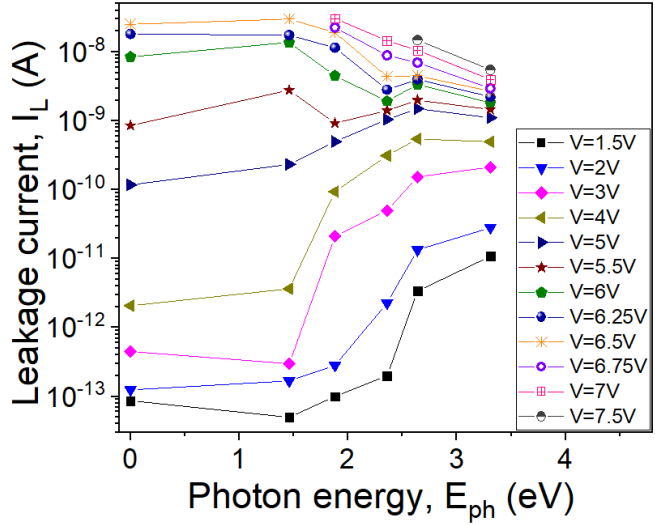

*Figure 6. The leakage current as a function of the photon energy. The parameter is the applied voltage. Note that the sample voltage, $V_S$, is lower than the applied value, since $V_s=V-IR_s$.*

    The reversed photoeffect is observed at bias voltages $V_S > 5.5$ V. A convenient way to present the effect is to plot the leakage current at a function of the photon energy, with the parameter being the applied voltage (Fig.6). The plots show a dichotomy: the leakage increases with the photon energy if the bias voltage is set to less than 5.5 V, while at higher voltages the current decreases with the photon energy.

    Motivated by previous observations of branching electronic patterns[20], as well as previous theoretical results suggesting that electrons penetrate in a disordered medium is a form analogous to a sandpile, characterized by a critical slope, we propose that electrons penetrate into the dielectric of nanocapacitors in a form of a branching sandpile structure schematically shown in Fig.7. The main feature of this EBSS is an increase of the current flowing through the dielectric due to (i) a reduction of the effective thickness of the dielectric and to (ii) an increase of

concentration of the electric field lines at the tips of the branched electronic pattern. The observed reversed photoeffect can then be naturally understood as a photon assisted disintegration of the EBSS[21].

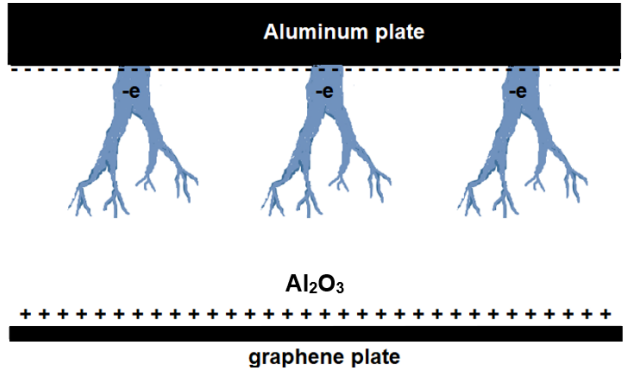

*Figure 7. Electronic branching "sandpile" structure (EBSS) developing in the insulating Al$_2$O$_3$ layer of the nanocapacitor. The tree-like structure illustrates penetrating electrons.*

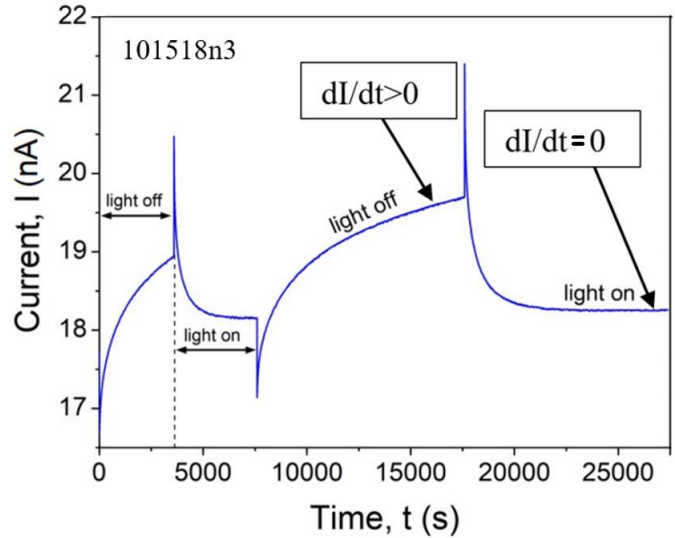

*Figure 8. Current through a representative nanocapacitor as a function of time at different regime: light off and light on. The voltage applied to the sample is 7 V. The insulating layer thickness is 10 nm. The light wavelength is 470 nm. When the light if OFF the EBSS is growing cause a slow but noticeable increase of the leakage. When the light is ON, the EBSS cannot grow and the current tends to a constant, i.e., dI/dt→0.*

To test the model, we perform time-dependent measurements of the current under alternating light exposure and darkness (Fig.8). In this experiment the applied voltage is fixed, $V$=7 V. When the light is off the current is steadily and slowly increasing, roughly logarithmically in time. This current increase represents the process of the EBSS growth, i.e., the electrons leak into the dielectric layer and form a charged structure inside. As the light is switched on, the EBSS gets destroyed and the current shows a sharp peak (the first such peak is marked by the dashed line). Then the current drops to a lower than the "light=off" level, because under the light exposure the EBSS cannot form. We note that without EBSS, i.e., when the light is on, the current approaches a steady state. This is different from the "light off" condition, in which the current tends to grow at all times, slowly but without any signs of approaching a saturation point.

Note that the behavior outlined above is different from the light doping experiments. There the light doping causes an increase of the conductivity as well as slow transient processes. In our case light exposure leads to a reduction of the current to a value that appears to be constant in time. The areas where EBSS grow (no light, $dI/dt>0$) and where EBSS are absent (light is on, $dI/dt=0$) we show by arrows in Fig.8.

DISCUSSION

The electronic pattern, referred to as EBSS here, is formed due to the fluctuations of the potential that can have local minima, i.e., electronic "traps" inside the dielectric layer. As we increase the bias voltage, the density of the electrons on the cathode (the negative plate of the capacitor) increases and, due to mutual Coulomb repulsion, some of the electrons begin to penetrate in the adjacent layer of the dielectric. In the dielectric, they fall into traps and become static, until their density and mutual repulsion is so high that some percentage of them begin to penetrate deeper into the dielectric. Thus, a "sandpile" pattern forms, where the electron density is high near the cathode and decreases linearly into the dielectric till it reaches zero at certain point. The boundary of the sandpile structure can have a branching shape since the electrons tend to leak into the insulator along easier paths and might form charged "rivers" or "branches"[20].

The electrons participating in the sandpile structure are not static but contribute to the leakage current, essentially making it larger (Fig.8). The EBSS grow approximately logarithmically with time (Fig.8), facilitating the corresponding growth of the leakages.

It might seem surprising that the electrons participating in the sandpile (EBSS) are not completely static. The reason for the current is as follows. The EBSS is characterized by a linear or at least a monotonically decreasing change density. This gradient of the charge density provides a consistent force which propels electrons deeper into the dielectric (like in an actual sandpile the sand grains can flow along the slope of the sandpile). If there were no thermal and/or quantum fluctuations, the sandpile boundary would propagate to a certain level inside the dielectric and then stop. Yet the electrons located at or near the boundary of the sandpile pattern can tunnel to the positive plate of the capacitor or, alternatively, they can escape from EBSS by the field emission process or by hopping, which depends on the applied voltage and the temperature. As some electrons leave EBSS, new ones come from the cathode to maintain a fixed charge density profile. Hence the electric current continues to flow "over the tilted slope" of the electronic sandpile, similarly to Ref. 14. So, the branches of the EBSS act as "launching paths" for the electrons, helping them to move closer to the anode and then either tunnel or hop.

The electronic sandpile pattern, EBSS, represent a locally elevated charge density. If an external perturbation provides sufficient energy, the electrons can escape EBSS and move to the anode and, thus contribute to the measured current. The typical depth of traps is significantly smaller than typical photon energies in our experiment. Therefore, each photon is capable to provide enough energy to one or more electrons participating in EBSS to release the electrons from the trap. As the light is applied the entire EBSS disintegrates, and the electrons simply gain energy from photons and tunnel to another electrode rapidly. That is why we observe current peak in Fig.8 as the light is turned on.

CONCLUSIONS

In summary, we present an optically transparent Al-Al$_2$O$_3$-graphene nanocapacitor suitable for conducting experiments related to the photoeffect and electron-photon interactions in general. The capacitor exhibits a field-emission current through the insulator at sufficiently large voltages, namely a fraction of a giga-Volt per meter. We find that the field emission effect is strongly suppressed when negative polarity is applied to the graphene plate of the capacitor. Our main finding is the reversed photoeffect, in which the leakage current is reduced by subjecting the device to electromagnetic radiation, namely visible light. The reversed photoeffect becomes greater when the photon energy is increased. We propose a qualitative explanation to this effect in terms of an electron charge self-organized structure, which forms in the insulating layer at high bias voltages. This structure helps electrons to tunnel through the insulator, but the applied influx of external photons dissolves the electronic pattern and thus suppresses the leakage current in the nanocapacitor.

METHODS

**Sample fabrication process.**

Metal-insulator-graphene capacitors were fabricated on glass substrates. The bottom layer is aluminum film deposited by mean of electron-beam evaporation in a chamber with the base pressure ~10$^{-9}$ Torr (AJA). After the aluminum deposition the samples were transferred (in air) into an atomic layer deposition (ALD) system where the whole surface was homogeneously covered by alumina using trimethylaluminum (TMA) and water vapor precursors at 80°C. The last step of the sample fabrication process was the deposition of a CVD (chemical vapor deposition) graphene (Trivial Transfer Graphene from ACS Material). It was transferred on the surface using the standard procedure provided by the manufacturer.

**The experimental circuit.**

The experimental circuit contains three main elements connected in series, namely a nanocapacitor, a series resistors $R_s$, a voltage source (Fig.1, insert). The current in the circuit is measured using Keithley 6517B ammeter, which provides a sub-pA current resolution. This device is also equipped with an adjustable voltage source, which provides voltage biasing, $V$. The standard resistor, $R_s =100$ MΩ, serves to limit the current in case of a breakdown. The sample was placed into a metal enclosure (Faraday cage). The Faraday cage was equipped with a photodiode for the purpose of light exposure. The measurements were controlled using LabView software.

ACKNOWLEDGMENTS

This work was supported by the Air Force grant AF FA9453-18-1-0004.